\documentclass[11pt,twoside]{article}
\usepackage{verbatim}
\usepackage{amssymb}
\usepackage{amsfonts}
\usepackage{amsmath}
\usepackage{pifont}
\usepackage[spanish,english]{babel}
\usepackage[pdftex]{graphicx}
\usepackage{epsfig}
\usepackage{bm}
\usepackage{longtable}
\usepackage{fancyhdr}
\usepackage{wrapfig}
\usepackage{cite}
\usepackage[pdftex]{hyperref}

\begin{document}

\fancypagestyle{plain}{%
\fancyhf{}%
\fancyhead[LO, RE]{XXXVIII International Symposium on Physics in Collision, \\ Bogot\'a, Colombia, 11-15 September 2018}}

\fancyhead{}%
\fancyhead[LO, RE]{XXXVIII International Symposium on Physics in Collision, \\ Bogot\'a, Colombia, 11-15 September 2018}

\title{Ultra-High Energy Neutrinos}
\author{James Madsen$\thanks{%
e-mail: james.madsen@uwrf.edu}$~ for the IceCube Collaboration\footnote{\protect\url{http://icecube.wisc.edu}}, \\ Physics Department, University of Wisconsin-River Falls, \\ River Falls, WI USA}
\date{}
\maketitle

\begin{abstract}
Ultra-high energy neutrinos hold promise as cosmic messengers to advance the understanding of extreme astrophysical objects and environments as well as possible probes for discovering new physics.  This proceeding describes the motivation for measuring high energy neutrinos.  A short summary of the mechanisms for producing high energy neutrinos is provided along with an overview of current and proposed modes of detection.  The science reach of the field is also briefly surveyed. As an example of the potential of neutrinos as cosmic messengers, the recent results from an IceCube Collaboration real-time high energy neutrino alert and subsequent search of archival data are described.     
\end{abstract}

\section{Introduction}

Astrophysics has seen an incredible expansion in the last century or so with new ways of viewing the cosmos revealing extreme objects and environments far exceeding what can be reproduced even in the most ambitious particle accelerator facility.  High energy astrophysics is an exciting field fueled by advances in the capabilities of observatories across the electromagnetic spectrum and in the size and scope of cosmic ray experiments.  Inherently interesting in their own right, astrophysical phenomena also provide a natural source to extend particle physics studies one or more orders of magnitude higher in energy. Cosmic rays and gravity waves were covered in separate talks at this meeting, as were neutrino results from reactors, accelerators, and neutrino-less double beta-decay experiments.  Lower energy neutrino topics, such as searches for sterile neutrinos, solar neutrinos, and coherent neutrino scattering were also presented separately.  Here the focus is on ultra-high energy (UHE) neutrinos of order 1 PeV and above, as well as the recent IceCube results associated with blazar TXS 0506+056.  

Neutrinos as cosmic messengers have many appealing characteristics. They are electrically neutral, traveling in straight lines from their source unaffected by magnetic fields. They interact only through the weak force, so they are mostly unaffected by intervening matter. They also track nuclear processes and can be used to study beyond standard model physics. The paths of nuclear cosmic rays, primarily protons but also heavier nuclei, are altered by magnetic fields so the detected direction does not point back to their source.  Extremely high energy cosmic rays will interact with the cosmic microwave background (CMB).  This is known as the GZK effect \cite{PhysRevLett.16.748, osti_6066749}; it limits the horizon for seeing protons with energies of 40 EeV and above to less than 100 Mpc.  The universe is also opaque to very high energy gamma rays due to interactions with the CMB.  The highest energy gamma rays detected are of order a hundred TeV, and it is likely gamma rays with energies of order PeV or more would have to be galactic in origin. Neutrinos seem to be the only currently known UHE messenger  capable of reaching the Earth from throughout the entire non-thermal universe.  

\section{UHE Neutrinos}
The compelling case for pursuing UHE neutrinos is tempered by daunting experimental challenges.  From the beginning, it was suggested that neutrino astronomy would require a detector a cubic kilometer in volume to have a hope of collecting enough neutrinos of astrophysical origin to study the cosmos. The event rate is determined by geometric factors such as the cross sectional area and angular acceptance, the detection efficiency which is dependent on both hardware and software design, the neutrino flux, and the neutrino cross section.  The production, detection, and science reach of UHE neutrinos are covered in this section.   
\subsection{Production}
\begin{wrapfigure} [16] {R}{.25\textwidth}
\vspace{-.3in}
    \begin{center}
        \includegraphics[width=.23\textwidth]{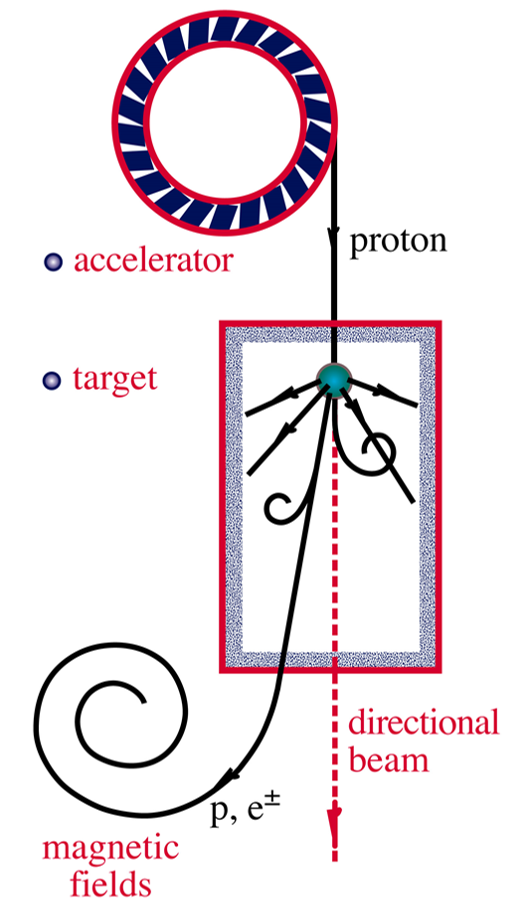}
  \end{center}
  \vspace{-.3in}
\caption{ Generic conditions for producing a neutrino beam.}
\label{fig:NP}
\end{wrapfigure}

A high-level representation of neutrino production is shown in Fig.~\ref{fig:NP}.  An accelerator produces high energy charged particles, shown in the figure as protons but for astrophysical accelerators heavier nuclei may also participate. 
The hadronic beam interacts with a target (a beam dump for particle accelerators), radiation or dust, that could be within the same environment where the acceleration took place, or somewhere along the cosmic journey.  The interaction produces secondary mesons; the charged mesons decay producing neutrinos and neutral mesons decay producing gamma rays. For example, neutral pions decay as $\pi^0\to\gamma+\gamma$ and create a flux of high-energy gamma rays; the charged pions decay into three high-energy neutrinos ($\nu$) and anti-neutrinos ($\bar\nu$) via the decay chain $\pi^+\to\mu^++\nu_\mu$ followed by $\mu^+\to e^++\bar\nu_\mu +\nu_e$, and the charge-conjugate process.

The flux of nuclear cosmic rays provides a test beam of neutrinos.  Cosmic rays that interact with nuclei in the Earth's atmosphere produce pions and kaons that decay giving a measured flux of neutrinos that agrees with model predictions.  The flux of these atmospheric neutrinos becomes very low at energies above $\sim$100 TeV as the probability increases that the charged mesons interact before they have a chance to decay.  A predicted prompt component at $\sim$100 TeV from the decay of charm mesons has yet to be seen.  A flux of astrophysical neutrinos with a harder energy spectrum compared to atmospheric neutrinos extending from below 100 TeV to about 10 PeV has been observed.  It will be discussed in section 3. A flux of extremely high energy neutrinos is predicted to result from off-source GZK interactions of extremely high energy cosmic rays with the CMB.  This cosmogenic neutrino flux is significantly smaller if the extremely high energy cosmic rays are composed of heavier nuclei rather than being predominately protons.  

\subsection{Detection}

\begin{wrapfigure} [11] {R}{.4\textwidth}
\vspace{-.3in}
    \begin{center}
        \includegraphics[width=.38\textwidth]{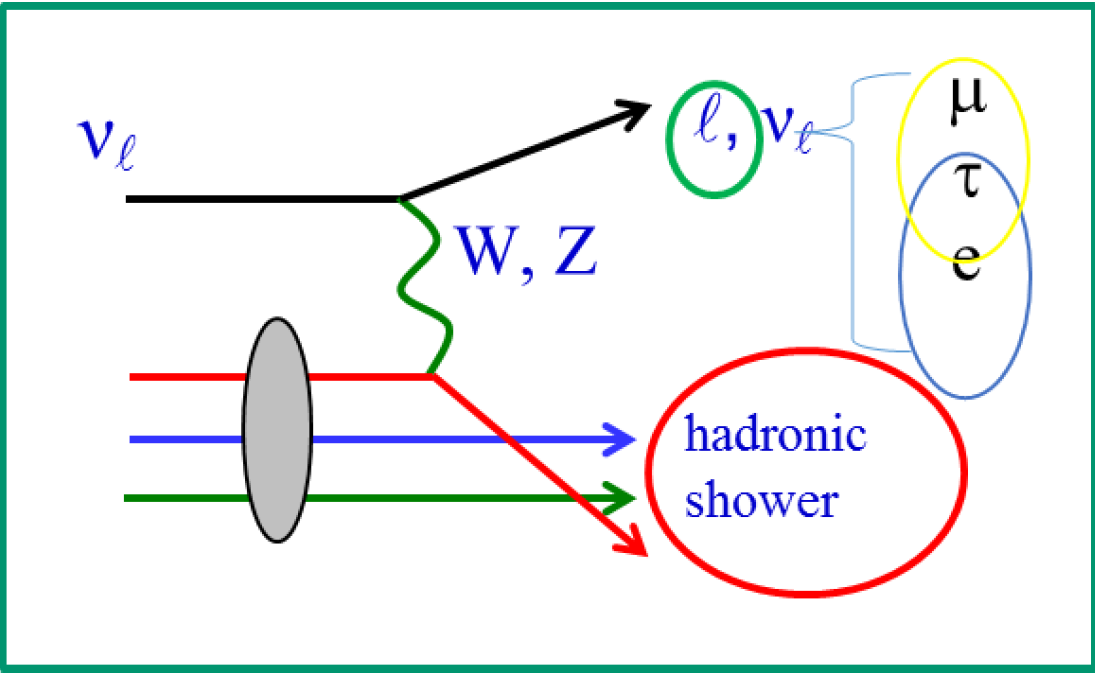}
  \end{center}
  \vspace{-.3in}
\caption{Neutrino interactions.}
\label{fig:NI}
\end{wrapfigure}

There are three flavors of neutrinos (l=muon, electron, or tau) which can interact by exchanging a W boson (charged current interaction) or a Z boson (neutral current interaction) as indicated in Fig.~\ref{fig:NI}. Aside from the Glashow resonance at about 6.3 PeV which involves an electron anti-neutrino~\cite{PhysRev.118.316}, there is not currently a way to experimentally distinguish between a high energy neutrino and anti-neutrino event. Knowing the flavor of the neutrino provides valuable information since astrophysical models make testable predictions of the relative numbers of each type that should be seen.  The detection process relies on the production of high energy charged particles traveling faster than the light speed in the medium producing Cherenkov radiation (near ultraviolet to visible), or the Askaryan effect (radio pulses for E$>$1 PeV), or from fluorescence in the atmosphere.  There are currently three approaches in routine use to search for high energy neutrinos.

\begin{figure*}[ht]
\includegraphics[width=.9\textwidth]{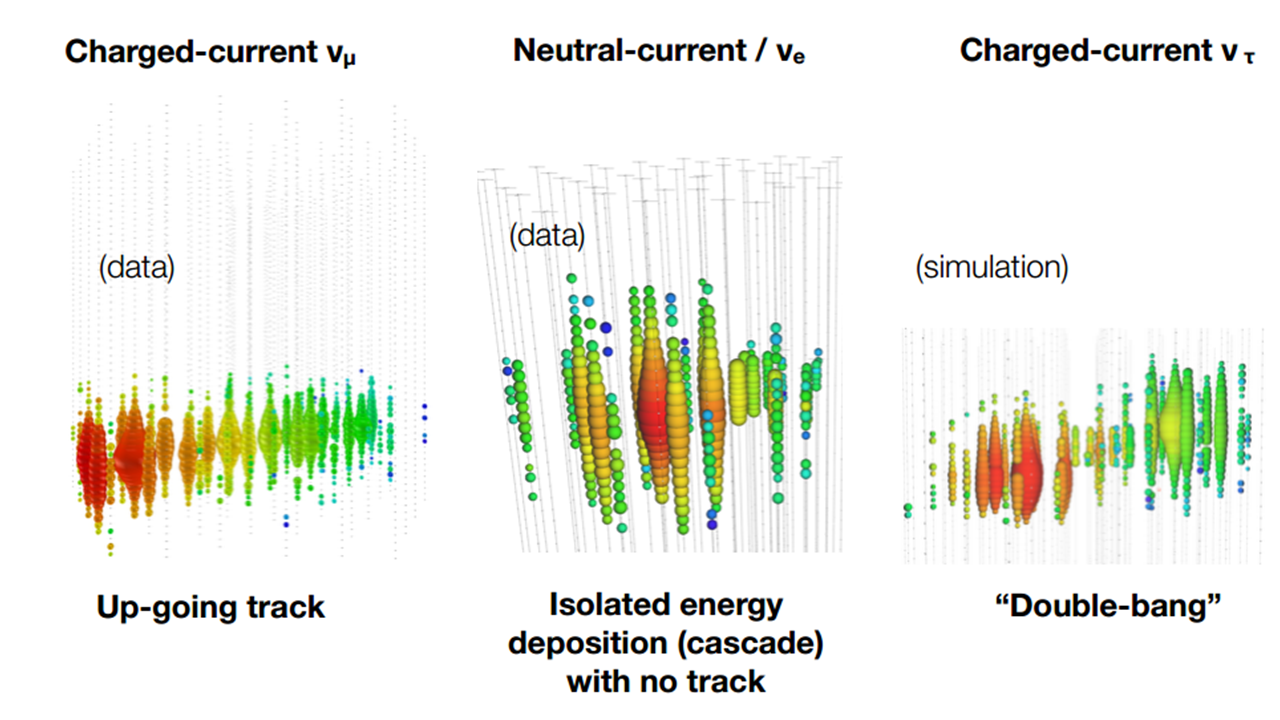}
\vspace{-.25in}
\caption{ Event classes for the IceCube Neutrino Observatory.  Each sensor is represented by a dot which is colored in with a circle whose diameter is proportional to the number of detected photons.  The color shows the time sequence, red first through blue last.}
\label{fig:ICSIG}
\end{figure*}

 An optically transparent medium, water or ice, can be instrumented with a grid of photosensors. The science goals determine the optimal configuration: smaller more densely instrumented arrays for low energy phenomena and larger more sparsely instrumented volumes for high energies.  This approach is used by Antares(antares.in2p3.fr) and KM3NeT(www.km3net.org) in the Mediterranean Sea, GVD(baikalgvd.jinr.ru) in Lake Baikal, and IceCube(icecube.wisc.edu) at the South Pole. The three neutral current interactions produce only a hadronic shower that is difficult to distinguish from the electron charged current interaction.  The muon neutrino charged current interaction produces a high energy muon with an identifiable track of light.  Tau charged current interactions produce two showers, one for the initial interaction that also gives a high energy tau.  The tau travels an average distance of 50m/PeV before decaying, resulting in a second shower. For sufficiently high energies, the two showers can be resolved.  Examples of the event types seen with the IceCube Neutrino Observatory are shown in Fig.~\ref{fig:ICSIG}.
 
 An alternative to detecting light with photosensors is to use antennas embedded in a radio transparent medium such as the ice in Antarctica or Greenland.  The advantage over optical detection is that the radio antennas are cheaper, and that the radio signals travel five to ten further than visible light.  This means large volumes of ice can instrumented at the required spacing for less cost.  The trade off is a much higher energy threshold, at least 1 PeV and perhaps an order of magnitude or two higher than that.  Two examples of this approach with test stations in place are the Askaryan Radio Array (ARA, ara.wipac.wisc.edu/home) at the South Pole and The Antarctic Ross Ice-Shelf ANtenna Neutrino Array (ARIANNA, arianna.ps.uci.edu).  
 
\begin{figure*}[ht]
\includegraphics[width=.5\textwidth]{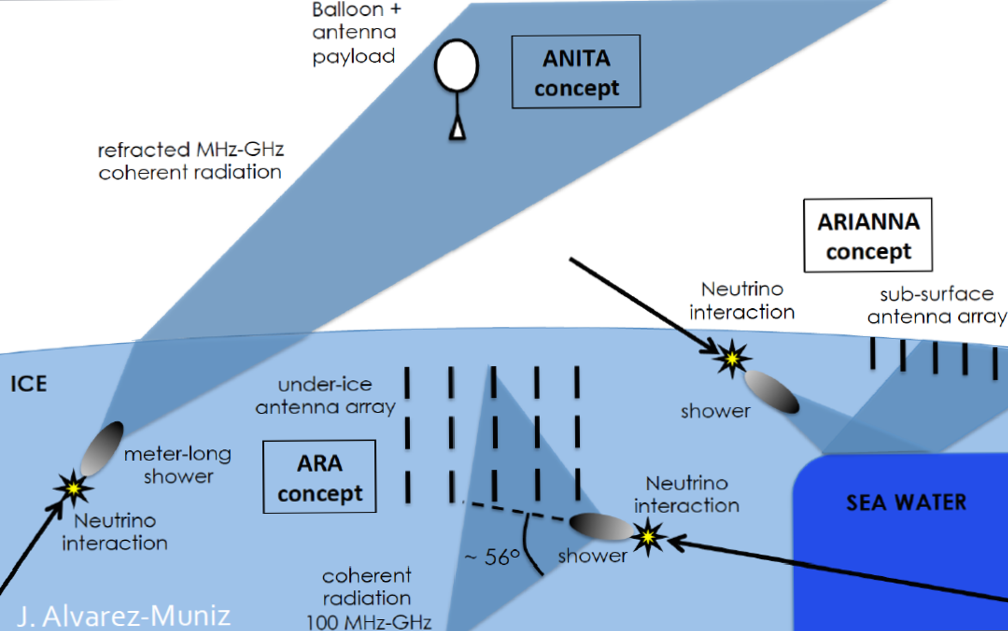}
\includegraphics[width=.5\textwidth]{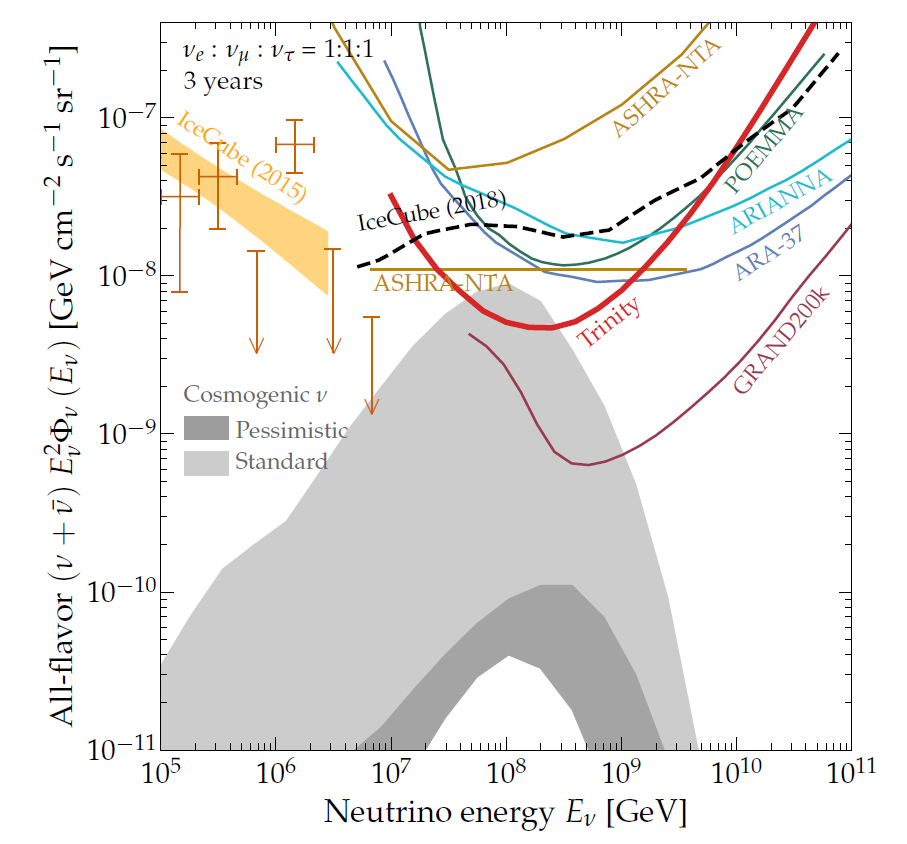}
\vspace{-.25in}
\caption{Examples of current projects to detect neutrino interactions using radio detectors(left)~\cite{Alvarez-Muniz:2017jxp} and their calculated sensitivities compared to cosmogenic neutrino flux models(right)~\cite{Otte:2018uxj}.}
\label{fig:Radio}
\end{figure*}

 Radio waves will also propagate with little loss through space and the atmosphere. The ANtarctic Impulsive Transient Antenna (ANITA, www.phys.hawaii.edu/\textasciitilde anita/) looks for the radio signal from neutrino interactions via balloon-borne receivers that fly at an altitude of about 35 kilometers, viewing about one million cubic kilometers of ice for about one month. The threshold energy is of order 10$^{18}$ eV, and no definitive neutrino events have been observed to date in four flights.  The detection concepts for ANITA, ARA, and ARIANNA are shown on the left in Fig.~\ref{fig:Radio}. Projects have also been proposed to look for radio waves with densely instrumented surface radio antenna arrays on the Earth; one example is the Square Kilomter Array (SKA)~\cite{refId0}. 
 
 Existing cosmic ray air shower experiments such as Auger (www.auger.org) or the Telescope Array (www.telescopearray.org) can also look for distinctive signals from ultra-high energy neutrino interactions that could occur much deeper in the atmosphere than nuclear cosmic ray interaction.  The resulting air shower can be imaged optically with imaging telescopes to see Cherenkov and Fluorescence light or by a sufficient large and densely instrumented radio array.  Neutrinos have a low probability of interacting in the relatively low density atmosphere and no events have been seen to date with this approach.  Horizontal neutrinos pass though more atmosphere, increasing the chance of interaction.  All flavors and both charged and neutral current interactions should be visible.  
 
 Earth skimming techniques rely on a tau neutrino charged current interaction occurring during a traverse through a small chord through the Earth's surface or while passing passing through a mountain. The much greater density increases the interaction probability but the shower from the first interaction in the Earth is not visible. The shower from the tau decay can be seen if it occurs in the Earth's atmosphere. For more details on the approach see for example a recent paper on the proposed Trinity project~\cite{Otte:2018uxj} that would use imaging telescopes or the white paper on the proposed Giant Radio Array for Neutrino Detection (GRAND) project~\cite{Alvarez-Muniz:2018bhp}. A summary of the expected sensitivities of several ultra-high energy neutrino telescopes compared to standard (proton) and pessimistic models (heavier nuclei) of the cosmogenic flux are shown on the right in Fig.~\ref{fig:Radio}~\cite{Otte:2018uxj}.

 \subsection{Science Reach}
 \begin{wrapfigure} [21] {R}{.7\textwidth}
 \vspace{-.4in}
    \begin{center}
        \includegraphics[width=.68\textwidth]{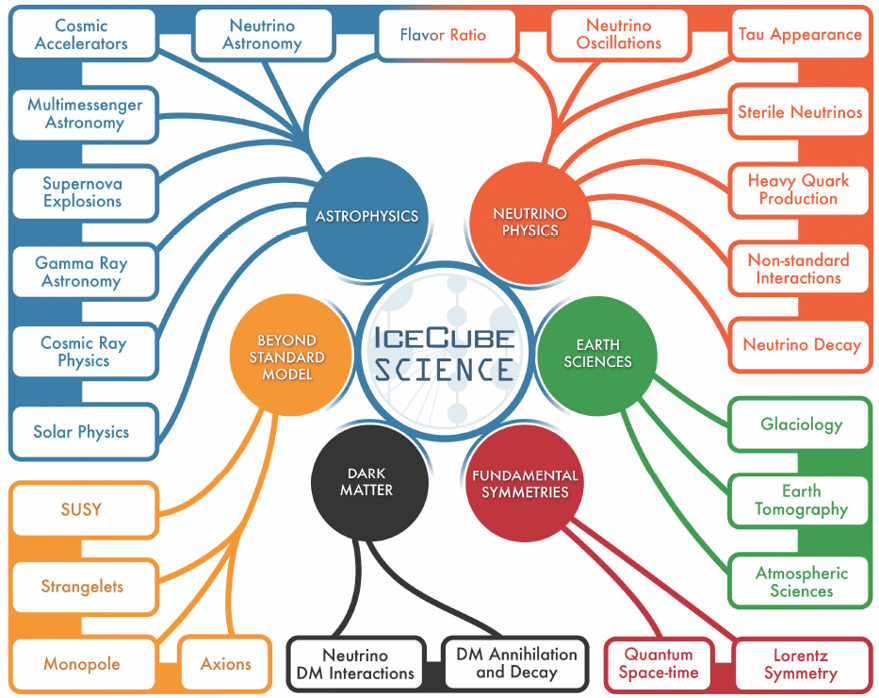}
  \end{center}
  \vspace{-.25in}
\caption{Neutrino observatories contribute to a wide range of topics.}
\label{fig:SR}
\end{wrapfigure}Neutrino astronomy may be a main motivator for the construction of facilities to detect high and ultra-high energy neutrinos, but it is not the only one. There is a remarkably wide range of physics that can be probed at energies at and significantly above those that can be achieved with particle accelerators.   A graphic showing the science reach of the IceCube Neutrino Observatory is shown in Fig.~\ref{fig:SR} and a review of IceCube particle physics has recently been published \cite{Ahlers2018}. 

The science capabilities for water-based neutrino observatories with similar energy threshold and total instrumented volume would be the same,  with glaciology replaced by studies associated with fresh or salt water environments. For the much higher energy threshold experiments that are targeting cosmogenic neutrinos, the range of science beyond probing cosmic ray interactions at the highest energies and doing neutrino astronomy, is more limited.  However, they would provide unique opportunities to extend studies of neutrino physics to ultra-high energies, possibly revealing new physics.   

\section{IceCube}
\begin{wrapfigure} [19] {R}{.55\textwidth}
\vspace{-.3in}
    \begin{center}
        \includegraphics[width=.53\textwidth]{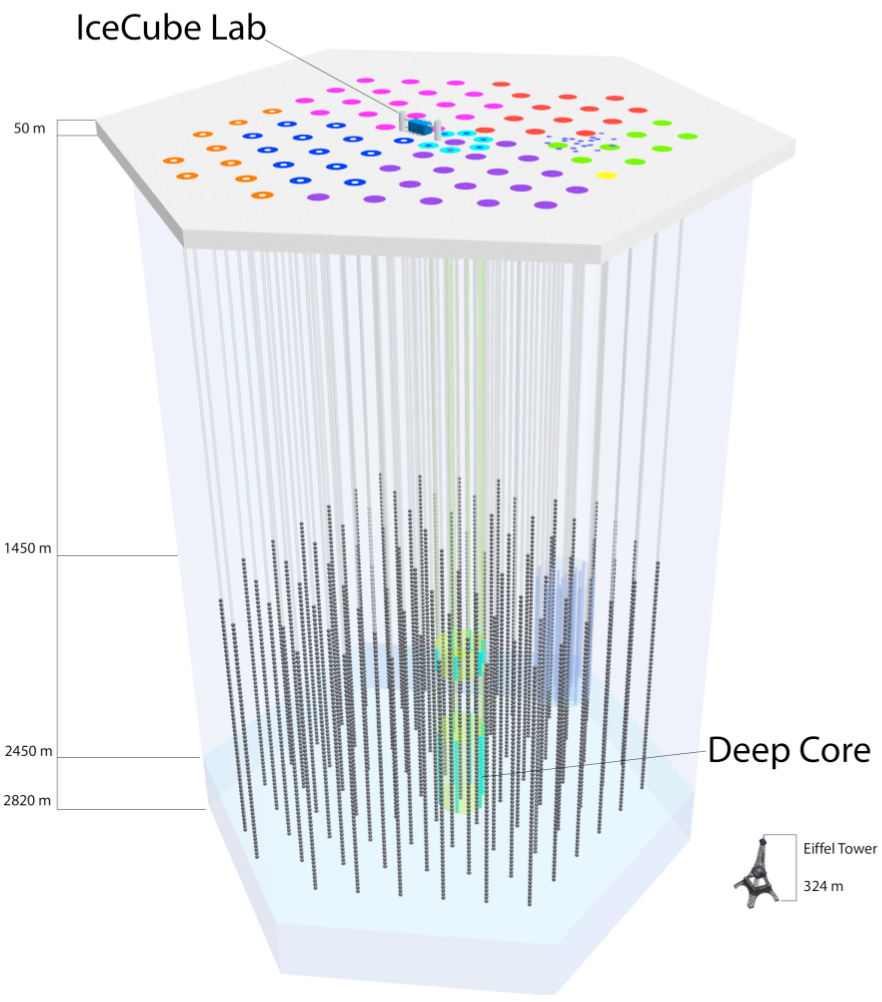}
  \end{center}
  \vspace{-.3in}
\caption{The IceCube Neutrino Observatory.}
\label{fig:IC}
\end{wrapfigure}

The IceCube Neutrino Observatory (IceCube) consists of a cubic kilometer of instrumented ice starting 1450 meters below the surface at the South Pole as shown in Fig.~\ref{fig:IC}.  It was constructed over seven seasons using a hot water drill to produce 60 cm diameter holes 2450 deep that were then instrumented with 60 light sensors.  Each light sensor, known as a digital optical module (DOM), has an internal clock and data acquisition system.  Power from the surface and data communication to the DOMs is provided by a cable; the cable with the 60 DOMs is referred to as a string.  Seventy-eight strings are spaced on a nominal 125 meter hexagonal grid, with the 60 DOMs 17 meters apart between 1450 and 2450 meters below the surface.  The central region, DeepCore, has 8 more closely spaced strings with most DOMs in the lower half of the ice enabling lower energy neutrinos to be measured.  A surface array, IceTop, consists of 81 stations nominally on the same 125 meter grid as the strings. 

The dominate signal seen by IceCube is from muons produced in cosmic ray interactions in the atmosphere in the southern sky.  This results in about $10^{11}$ events per year. About one in a million of these events can be identified as neutrinos (about $10^{5}$ per year) either by their trajectory from the northern hemisphere, where the Earth filters cosmic ray muons, or by looking for events that begin inside the array of DOMs.  The latter approach allows neutrino events to be seen from all directions.  Most of the measured neutrinos are produced by cosmic ray interactions in the atmosphere but about 10 to 20 times per year a neutrino event is identified with an energy of 100 TeV or more.  The higher the energy, the less likely it is to be from an atmospheric interaction and the more likely it is from an astrophysical source.  
\subsection{Neutrino Spectrum} 
The neutrino flux above 1 TeV measured by IceCube is shown on the left in Fig.~\ref{fig:IFXCC}.  The flux attributed to muon neutrinos and anti-neutrinos from cosmic ray interactions in the Earth's atmosphere from pion and kaon decay is shown in blue.  No evidence is seen for a prompt atmospheric neutrino flux from charm meson decays so a limit is shown in green. The salmon color shows the astrophysical neutrino flux determined from muon neutrino events originating in the northern hemisphere~\cite{Aartsen:2016xlq}. The data points are the corresponding flux determined by an independent analysis that looked for high energy starting events (HESE) that began inside the detector. The agreement is good between the two results for energies above 200 TeV with a single power fit to the astrophysical muon neutrinos flux having an exponent of 2.19 $\pm$ 0.1~\cite{Niederhausen:2017mjk}.

\subsection{Neutrino Cross Section}
\begin{figure*}[ht]
\includegraphics[width=.5\textwidth]{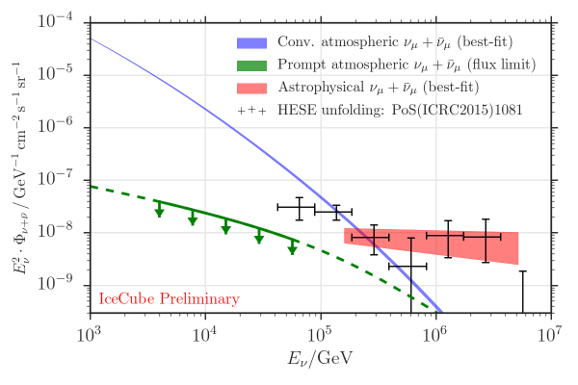}
\includegraphics[width=.5\textwidth]{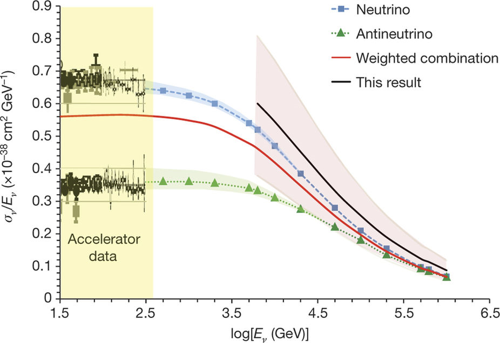}
\vspace{-.3in}
\caption{The measured IceCube neutrino flux (left) and neutrino cross section (right).}
\label{fig:IFXCC}
\end{figure*}

Well-reconstructed muon neutrino events can be used to determine the neutrino cross section at energies more than order of magnitude higher than previous measurements.  For energies above 40 TeV, the adsorption length is less than one Earth diameter, and neutrinos that pass through the center of the Earth do not make it to the IceCube array.  Events with  larger zenith angles traverse less of the Earth, so neutrino cutoff energy increases with zenith angle. The neutrino cross section was determined by measuring the attenuation of 10,784 IceCube neutrino events with energies between 6.8 TeV and 980 TeV as a function of energy and zenith angle, see Fig.~\ref{fig:IFXCC} (right). The result was 1.30 $\pm$$^{+0.21}_{-0.19}$(stat.)$\pm$$^{+0.39}_{-0.43}$(syst.)times the standard model predictions including charged and neutral current interactions, and is statistically consistent with the standard model prediction\cite{Aartsen:2017kpd}. 

\subsection{Online Alerts and the IceCube event 170922A}
The nominal event rate for IceCube is 2.7 kHz.  Sufficient processing power is available at the South Pole to do a fast first fit of the data to look for likely astrophysical neutrino events.  A second astrophysical filter is based on an analysis that looks for extremely high energy (EHE) through going tracks events. The EHE energy criteria for the online alert was lowered to .5 PeV to have a higher capture probability although at a lower signal purity. Events that satisfy the HESE criteria and are track-like rather than a cascade or that pass the EHE filter are processed with a second relatively fast fitting algorithm, and an automated alert is sent out typically in less than a minute.  The public online channel that was activated in April 2016 and a complete description of the alert filters and processes that result in a signal efficiency of 30-50\% is available. \cite{Aartsen:2016lmt}

\begin{figure*}[ht]
\includegraphics[width=.5\textwidth]{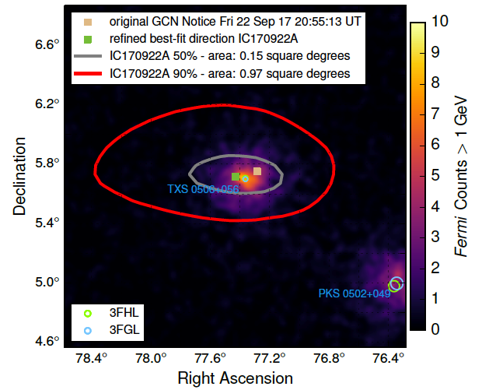}
\includegraphics[width=.5\textwidth]{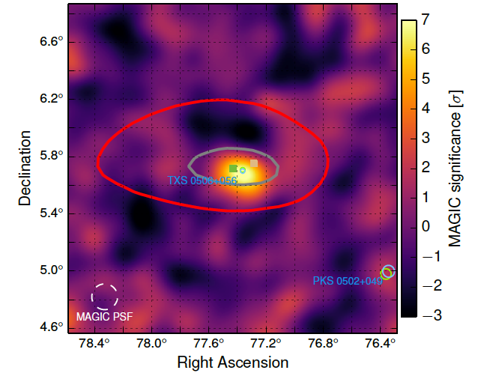}
\vspace{-.3in}
\caption{The observations for Fermi (left) and Magic (right) gamma-ray telescopes showing the IceCube reconstruction error contours for event 170922A.}
\label{fig:FM}
\end{figure*}
On September 22, 2017 at 20:54:30.43 UTC the IceCube EHE identified a well-reconstructed slightly upward going muon track event which subsequent analysis showed had an energy loss in the detector of loss 23.7 +/- 2.8 TeV corresponding to a most probable neutrino energy of 290 TeV~\cite{IceCube:2018dnn}.  Followup observations by Fermi-LAT reported a gamma ray flaring blazer TXS 0506+056 (ATel\#10791).  The blazer TXS 0506+056 was seen in further observations by the very high energy gamma-ray Magic telescope (Atel\#10817), with both observations well within the IceCube error contours as shown in Fig.~\ref{fig:FM}. Further observations determined that this blazer had a redshift of z = 0.3365 corresponding to a distance of about 4 billion lightyears~\cite{Paiano:2018qeq}.  The chance probability of a Fermi-IceCube coincident observation was calculated to be at the 3 sigma level.  A subsequent search of 9.5 years of IceCube archival data looking for events in the direction of TXS 0506+056 found 13 excess events compared to background expectation of 5 events between September 2014 and and March 2015~\cite{IceCube:2018cha}.  This number of events is inconsistent with the background only hypothesis at the 3.5 sigma level.  This calculation is independent of IceCube event 170922A which motivated this archival search.   

\section{Conclusions}
Nature has provided a beam of high energy neutrinos useful for astrophysics, particle physics, and tests of fundamental physics principles such as Lorentz invariance.  Building instruments capable of seeing statistically significant numbers of high energy neutrinos is a challenge.   However, significant progress has been made and there are multiple projects in operation that demonstrate the viability of the field.  The IceCube Neutrino Observatory is the first functioning cubic kilometer scale detector and there are plans for similiar scale facilities with photosensors deployed in water at sites in the northern hemisphere. IceCube has identified a flux of astrophysical neutrinos that extends up to at least 10 PeV. The neutrino alert from IceCube event 170922A that led to the multimessenger observations of blazar TXS 0506+056 is a major milestone for neutrino astronomy.  There are near term plans, the IceCube Upgrade, to significantly enhance IceCube's low energy capabilities and longer term plans for a IceCube Gen2 instrumented ice volume by an order of magnitude.  Measuring cosmogenic neutrinos produced when extremely high energy cosmic rays interact with CMB photons will require even larger instrumented volumes necessitating a different detector technology.  At present, the two most promising approaches are radio arrays and imaging telescopes to look for signals for tau neutrinos that interact in Earth skimming events or in mountains, and detecting the resulting air shower if the tau decays in the atmosphere.   

\section*{Acknowledgment}

Thanks to  the organizers for the invitation to present at the XXXVIII International Symposium on Physics in Collision, and the local hosts for the warm welcome and great hospitality during the conference.  Travel to the conference was made possible by professional development funds from the University of Wisconsin-River Falls.  

\bibliographystyle{ieeetr}
\bibliography{main}
\end{document}